\documentclass[twocolumn,showpacs,preprintnumbers,amsmath,amssymb,superscriptaddress]{revtex4}
\usepackage{graphicx}% Include figure files
\usepackage{dcolumn}% Align table columns on decimal point
\usepackage{bm}% bold math
\usepackage{epsfig}
\usepackage{CJK}
\usepackage{bm}

\begin{document}

\title{Topological Single Electron Pumping Assisted by Majorana Fermions}

\author{Qi-Feng Liang}
\affiliation{International Center for Materials Nanoarchitectornics (WPI-MANA)
National Institute for Materials Science, Tsukuba 305-0044, Japan}
\affiliation{Department of Physics, Shaoxing University, Shaoxing 312000, China}
\author{Zhi Wang}
\author{Xiao Hu}
\affiliation{International Center for Materials Nanoarchitectornics (WPI-MANA)
National Institute for Materials Science, Tsukuba 305-0044, Japan}
\pacs{03.67.Lx, 71.10.Pm, 74.45.+c, 74.78.Fk}
\date{\today}

\begin{abstract}
Single electron pumping based on the topological property of
Majorana fermions (MFs) is proposed. The setup consists of a quantum dot and four nano
topological superconductors (TSs) connected by constriction junctions, with an additional
vortex located in the loop of TSs. Operation is performed by gate voltages at
constriction junctions. Simulations with Bogloliubov-de Gennes equation demonstrate
successfully quantum protection during switching operation.
\end{abstract}

\maketitle

\section{Introduction}
In the electronics-based computers, computation
is performed based on motions of thousands of electrons, where the
accuracy of bit manipulation relies on the statistics of huge number of
electrons. If one can control electrons one by one at given instants, a
single electron bit may be achieved, which reduces the energy consumption
to the limit. This technology is also important for forming the metrology
triangle of Ohm's law: the quantum Hall effect discovered by von Klitzing
works as the conductance standard\cite{vonKlitzing}, and the Josephson
effect provides the voltage standard\cite{Josephson}, while realization of
precise current standard remains to be a challenge\cite{GiblinNatComm}.
\begin{figure}[tp]
  \includegraphics[width=7cm]{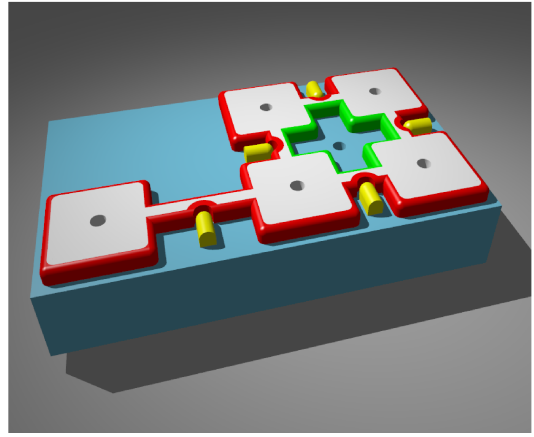}
  \caption{Schematic device
setup of a NOT gate for MF qubit. Finite square samples (called bricks)
of heterostructure of ferromagnetic insulator and spin-orbit coupling
semiconductor are positioned on the surface of an $s$-wave superconductor
with vortices (black dots) trapped in it. The four bricks (on the right
side) form the NOT gate. The edge MF of another brick (on the left side)
can be driven to circle the vortex at the center of the loop formed by the
four bricks by tuning gate voltages at the point-like constriction
junctions.}
 \label{Fig 1}
\end{figure}

Manipulation of individual electrons in condensed matters is not easy
since the wave length of electrons is comparable with their separation.
One way to transport single electrons is to use quantum
tunneling phenomenon. One needs to puncture the time for electrons to
tunnel through the barrier by introducing several quantum dots (QDs) in the
circuit, since quantum tunnelings normally take place randomly in
time\cite{LuNature,FujisawaAPL}.
Several schemes have been proposed for single electron pumping
so far, such as single electron transfer in metallic nanostructures based on the combination of Coulomb interaction
between electrons and their quantum tunneling trough an insulating barrier \cite{BiasCQ}, adiabatical charge pumping through a
quantum dot by driving two independent parameters of the system coherently, \textit{e.g.} phases of gate voltage and magnetic field\cite{PWBrouwer,AdiaCQ}, quantum Hall effect in Corbino disk geometry combined with a time dependent flux \cite{QHEpumping}.

In a topological superconductor characterized by zero-energy Majorana
fermions (MFs) as quasiparticle excitations, the ground state exhibits
degeneracy with respect to the parity of electron number
\cite{MooreRead91,ReadGreen,Ivanov,Kitaev01,Kitaev03,DasSarmaPRL05,superfluid,MSatoPRL09,Tewari_Index,FuKanePRL08,MFreturn,Akhmerov_Beenakker_prl09,SauDasSarmaPRL10,
LutchynDasSarma,OregvonOppen10,FisherNPhys,TQC_RMP}. This property
provides a new principle for single electron pumping with unprecedented
precision guaranteed by topology. In the present work we formulate a
protocol for this purpose.

Our setup consists of four topological superconductors (TSs)
each carrying a vortex and connected by constriction junctions; there
is an additional vortex inside the loop of TSs (see Fig.~1).
In order to reveal the topological property of the system, we first attach
another TS to the device. We show that the edge MF of this TS can be
driven in a controlled way around the loop of the four TSs by
switching on and off gate voltages at constriction junctions in the
designed sequence. After this process, the parity of this TS is flipped
since the edge MF acquires a $\pi$ phase due to the vortex at the center
of device. Therefore, one can consider this TS as a MF qubit, and the
device of four TSs as a NOT gate. We then demonstrate that the NOT gate
works for single electron pumping when a QD with Coulomb blockade
effect is attached when the energy of QD is adjusted appropriately. Numerical
simulations based on the time-dependent Bogloliubov-de-Gennes (TDBdG)
equation\cite{OurPaper} are performed, which confirms successfully the quantum
protection and phase coherence during the whole process typically of
several nano seconds.

The remaining part of this paper is organized as follows. We discuss in Sec.~II
the dynamics of an edge MF in a MF-qubit driven through the NOT gate.
Then we reveal the relation satisfied by the interactions in the NOT gate
in Sec.~III. Based on the property of NOT gate, we formulate in Sec.~IV the topological
single electron pumping by attaching a QD to the NOT gate, and give explicitly
the energy regime for QD. In Sec.~V, it is clarified that the function
of NOT gate can be understood as a quantum interference between two MFs. Discussions are presented in
Sec.~VI with a summary given in Sec.~VII.

\section{Not gate for MF-qubit}
A topological
superconducting state can be achieved in a heterostructure of $s$-wave
superconductor (s-SC), spin-orbit coupling semiconductor (SOSM), and
ferromagnetic insulator (FMI) with one vortex at the sample center, where
one MF appears at the vortex core and another MF at the sample edge
\cite{SauDasSarmaPRL10,OurPaper}. In the present device, four finite TSs
are positioned on a common $s$-SC substrate. The core MFs are stable and
do not participate in the phenomena discussed below (strictly speaking
there are exponentially small contributions, which can be neglected
safely), and thus will be omitted hereafter. The linear dimension of TSs
should be in the regime of tens to one hundred nano meters\cite{OurPaper}:
for too large TSs, the energy gap
between the zero-energy MF state and the lowest excited state becomes
very small, which limits the operation temperature, while for too small
TSs, core MF and edge MF interact with each other, which destroys the
MF ground state. Edge MFs interact with
neighboring ones through the constriction junctions when electron hoppings
are permitted. There is an additional vortex in the s-SC substrate inside
the loop formed by the four TSs, which governs the interactions among edge
MFs as will be revealed below. We attach a MF qubit to the above device as
schematically shown in Fig.~1.

The low-energy physics of the system can be described by the following
effective Hamiltonian,
\begin{equation}
\hat{H}_{\rm MF}= i\lambda_0(t)\Gamma_0\hat{\gamma}_0\hat{\gamma}_{1}
+i\sum_{j=1,4}\lambda_j(t)\Gamma_j\hat{\gamma}_j\hat{\gamma}_{j+1}
\label{Hmf}
\end{equation}
where $\hat{\gamma}_0$ denotes the edge MF at the qubit, and
$\hat{\gamma}_j$ with $1\le j\le 4$ denote those on the TSs in the NOT
gate and $\hat{\gamma_5}\equiv \hat{\gamma_1}$; time-dependent,
dimensionless factors $0\le \lambda_j(t)\le 1$ are introduced for the
switching process of constriction junctions, whereas $\Gamma_j$ are the MF
interactions when the gate voltage is off and thus two TSs are connected
fully.

At the initial stage, the switch configuration is described by the vector
$(\lambda_0, \lambda_1, \lambda_2, \lambda_3, \lambda_4)=(0,1,0,1,0)$
(see the left inset of Fig.~2).
Since the qubit is isolated, there is an edge MF ($\hat{\gamma_0}$)
localized at the qubit. In contrary, with $\lambda_1=\lambda_3=1$, both the
unified edge of TS(1) and TS(2) and that of TS(3) and TS(4) contain two
vortices, and thus there is no edge MF in the NOT gate. In the representation of
\ref{Hmf}, the MFs $\hat{\gamma_j}$ for $1\le j\le 4$ are fused to
finite energies due to the interactions.

Turning on the connection between the qubit and TS(1), namely
$\lambda_0=0\rightarrow 1$, the wavefunction of $\hat{\gamma_0}$ spreads
to the unified edge of the now connected qubit, TS(1) and TS(2), since it
contains three vortices. We then turn off the connection between TS1 and
TS2, namely $\lambda_1=1\rightarrow 0$. The wavefunction of edge MF
collapses totally on TS2 due to the topological property. After these two
switchings, the edge MF $\hat{\gamma_0}$ is transported completely to TS2.
Repeating this process, one can drive the edge MF $\hat{\gamma_0}$ through
the NOT gate in a clockwise way, and returns it back to the initial
position at the qubit, with the switching sequence
$(0,1,0,1,0) \mapsto (1,1,0,1,0)
\mapsto (1,0,0,1,0) \mapsto (1,0,1,1,0) \mapsto (1,0,1,0,0) \mapsto
(1,0,1,0,1) \mapsto (0,0,1,0,1)$. During this process the edge MF
feels the gauge field formed by the central vortex, and thus acquires a
phase of $\pi$ which makes $\hat{\gamma_0} \mapsto -\hat{\gamma_0}$. As
the result, the electronic parity of the qubit is flipped.

In order to confirm the function of the NOT gate, we perform numerical
calculations based on TDBdG equation. We first diagonalize the
tight-binding BdG hamiltonian \cite{OurPaper} of the system in the initial
stage,
\begin{align}
&\tilde{H}_0=-t_0\sum_{\bold{i},\bold{j},\sigma}
\hat{c}_{\bold{i}\sigma}^{\dag}\hat{c}_{\bold{j}\sigma}
- \mu \sum_{\bold{i},\sigma} \hat{c}_{\bold{i}\sigma}^{\dag}\hat{c}_{\bold{i}\sigma}+\sum_{\bold{i}}V_z (\hat{c}_{\bold{i}\uparrow}^{\dag}\hat{c}_{\bold{i}\uparrow}-\hat{c}_{\bold{i}\downarrow}^{\dag}\hat{c}_{\bold{i}\downarrow}) \nonumber\\
&+i t_{\alpha}
\sum_{\bold{i},\bold{\delta}}\left[\hat{c}_{\bold{i}+\bold{\delta}_x}^{\dag}\hat{\sigma}_y\hat{c}_{\bold{i}}-
\hat{c}^{\dag}_{\bold{i}+\bold{\delta}_y}\hat{\sigma}_x\hat{c}_{\bold{i}}+h.c.\right]+ \sum_{\bold{i}}\left[\Delta(\bold{i}) \hat{c}_{\bold{i}\uparrow}^{\dag}\hat{c}_{\bold{i}\downarrow}^{\dag}+h.c.\right],\\\nonumber
\end{align}
where both spin-conserved hopping $t_0$ and spin-flipped hopping
$t_{\alpha}$ are between nearest neighbors with $a$ the grid spacing, and
$V_z$ is the Zeemann energy. An edge MF state is obtained at the MF qubit
with the wave function $|\Psi(t=0)\rangle=|\phi_{\rm MF}\rangle$. Then we
modulate dynamically the hopping parameters at the constriction junctions, which
changes the MF interactions in \ref{Hmf} sequentially and drives the edge MF
$\hat{\gamma}_0$ \cite{OurPaper}. The
evolution of the wave function is obtained  by solving the TDBdG equation
$i\hbar\frac{\partial}{\partial t} |\Psi(t) \rangle= H|\Psi(t)\rangle$
based on the Chebyshev polynomials expansion\cite{tdbdg_1,tdbdg_2}. To
monitor the evolution of the edge MF, we project the wave function onto
the initial one, and evaluate the parameter $O(t)= \langle\phi_{\rm MF}
|\Psi(t)\rangle$. As can be seen from Fig.~2, $O(t)$ changes from positive
unity at the initial stage to negative unity at the final stage,
representing a sign change in the MF wave function. It is worth noticing
that the conservation of the function norm as seen in Fig.~2 confirms the
topological protection of the edge MF during the operating process.

\begin{figure}[t]
\includegraphics[width=6.5cm]{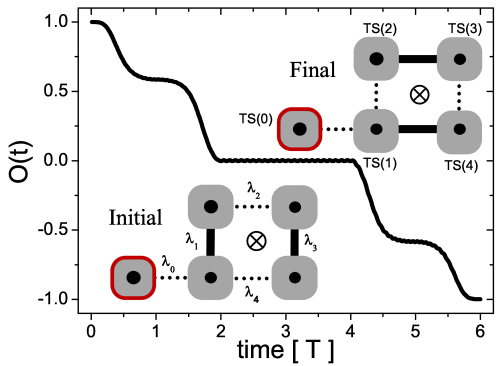}
\caption{(Color on line) Time
evolution of the MF wave function transported through the NOT gate
in terms of projections $O(t)$ to its initial wave function. The
time for one switching step is $T=4\times10^4\hbar/t_0$. The
two insets schematically show the initial and final stages, where the solid (dotted)
lines denote the on, $\lambda=1$ (off, $\lambda=0$) state of the constriction junctions between TSs.}
\end{figure}

\section{Parity flipping in NOT gate}
Since the parity of the
whole system is conserved upon application of gate voltage as well as
Cooper pair tunneling from SC substrate, the above switching operation
should reverse the parities of the MF qubit and NOT gate simultaneously.
In order to check the electronic parity of the NOT gate, we investigate
first the signs of MF interactions $\Gamma_j$ defined in Eq.~(1). As
revealed in the previous work\cite{OurPaper}, when $\hat{\gamma}_0$ is
transported to TS(2), it picks a sign sgn($\Gamma_0\Gamma_1$). Therefore,
the sign of edge MF $\hat{\gamma}_0$ after driven through the NOT gate is
given in terms of the interactions by

\begin{equation}
\hat{\gamma}_0 \Rightarrow \text{sgn}[\Gamma_0\Gamma_1\Gamma_2\Gamma_3\Gamma_4(-\Gamma_0)]\hat{\gamma}_0.
\end{equation}
The minus sign attached to $\Gamma_0$ is due to the
opposite motion of MF against the direction used to define the interaction
$\Gamma_0$ in \ref{Hmf}\cite{OurPaper}.

It is then clear that the sign reversal of $\hat{\gamma}_0$ implies
$\Gamma_1\Gamma_2\Gamma_3\Gamma_4>0$, a topological property generated by
the central vortex. Hereafter, we consider explicitly the
case where all interactions are positive since all possible configurations
of interaction signs can be transformed to each other by gauge
transformation. It is worth noting that the same sign constraint
$\prod_{j=1,2N} \Gamma_j>0$ and gauge choice are available for even number
of TSs, which will be used for discussions below.

We define two regular electronic states with the four MFs,
$\hat{d}_1^{\dag}=(\hat{\gamma}_1+i\hat{\gamma}_2)/2$ and
$\hat{d}_2^{\dag}=(\hat{\gamma}_3+i\hat{\gamma}_4)/2$, as always possible
even when the MFs are bounded and not free. We then rewrite the
Hamiltonian \ref{Hmf} in terms of the basis $|n_1n_2\rangle=\{|00\rangle,
|11\rangle, |10\rangle, |01\rangle\}$, with $n_i$ denoting the parity of
the electronic state,
\begin{equation}
H_{\rm MF}=\left[
           \begin{array}{cccc}
             \Gamma'_1+\Gamma'_3 & \Gamma'_2-\Gamma'_4     & 0                 & 0  \\
             \Gamma'_2-\Gamma'_4 & -(\Gamma'_1+\Gamma'_3) & 0                 & 0  \\
              0                & 0                     & \Gamma'_1-\Gamma'_3 & \Gamma'_2+\Gamma'_4 \\
              0                & 0                     & \Gamma'_2+\Gamma'_4 & -(\Gamma'_1-\Gamma'_3) \\
           \end{array}
         \right],
\label{He}
\end{equation}
where $\Gamma'_j\equiv \lambda_j(t)\Gamma_j$. The four eigen energies are
given by $ E_{1,2}= \pm
\sqrt{(\Gamma'_1+\Gamma'_3)^2+(\Gamma'_2-\Gamma'_4)^2}$ for even parity,
and $ E_{3,4}= \pm \sqrt{(\Gamma'_1-\Gamma'_3)^2+(\Gamma'_2+\Gamma'_4)^2}$
for odd parity. The switch configuration of the constriction junctions in
the NOT gate changes after the operation: $\lambda_1=\lambda_3=1$ and
$\lambda_2=\lambda_4=0$ at the initial stage, while
$\lambda_1=\lambda_3=0$ and $\lambda_2=\lambda_4=1$ at the final stage.
It is clear that the ground-state energy $E_{\rm g}=-(\Gamma_1+\Gamma_3)$ at the initial
stage is achieved in the even-parity subspace, while $E_{\rm
g}=-(\Gamma_2+\Gamma_4)$ at the final stage in the odd-parity subspace.
Therefore, the electronic parity of the NOT gate is reversed after the
operation of switching.
It is easy to see that this parity reversal is realized because
$\Gamma_1\Gamma_2\Gamma_3\Gamma_4>0$.

\section{Single electron pumping}
The NOT gate discussed
above can be used for single electron pumping when a QD is
attached (see Fig.~3(a)). The parity flipping specifies an odd number of electrons
transferring between the NOT gate and QD upon the switching operation
described above, whereas the Coulomb blockade effect on the QD limits the
charge transfer to a single electron.

\begin{figure}[t]
\includegraphics[width=6.5cm]{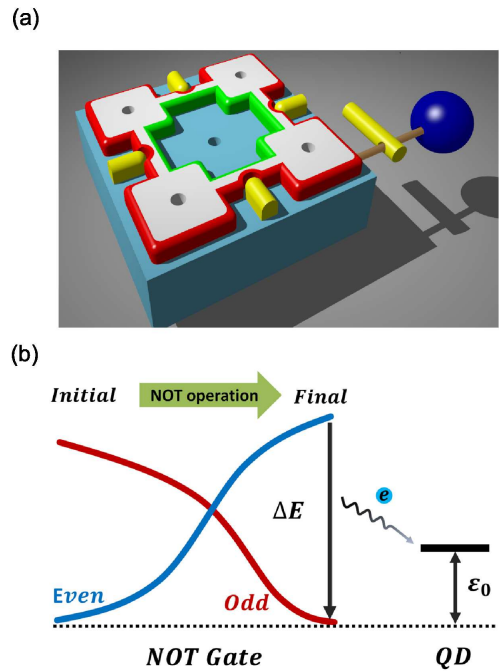} \caption{(a) Topological single electron
pumping realized by the NOT gate and QD. A gate voltage (yellow cylinder)
is used to tune the coupling between the QD and one TS of the NOT gate. (b) Working
mechanism of single electron pumping (see text). The two lowest energy levels (red and blue)
associated with opposite parities of the NOT gate cross each other upon the switching
operation, with $\Delta E$ the
energy difference at the final stage. When $\Delta E >\epsilon_0$,
the occupation energy of QD, an electron is emitted from the NOT gate to the QD
to lower the energy of whole system.}
\end{figure}

Let us first consider the case of a vacant QD with an energy level $\epsilon_0$ in
the initial stage (see Fig.~4(b)). The ground state of the total system
exhibits even parity, since the NOT gate is in a state of even parity.
From the parity conservation of the total system, there are two candidates
for the ground state at the final stage after switching operation discussed above: vacant QD with total energy
$-|\Gamma_2-\Gamma_4|$, and/or occupied QD with total energy
$-(\Gamma_2+\Gamma_4)+\epsilon_0$, both exhibit even parity. One electron
on the NOT gate is transferred to the QD upon the switching operation if
$-(\Gamma_2+\Gamma_4)+\epsilon_0<-|\Gamma_2-\Gamma_4|$, namely $\epsilon_0
<2\min\{\Gamma_2, \Gamma_4\}$. Physically the inequality means that if the
occupation energy of the QD is too large, electron will not jump to QD. In
the same way, one can figure out that when QD is occupied at the initial
stage, an electron is transferred to the NOT gate upon switching operation
if $-(\Gamma_2+\Gamma_4)<-|\Gamma_2-\Gamma_4|+\epsilon_0$, namely
$\epsilon_0>-2\min\{\Gamma_2, \Gamma_4\}$. The physical meaning of the
condition is also clear.

In order to perform single electron pumping starting from a switch
configuration opposite to the one discussed above, one has two similar
conditions on $\Gamma_1$ and $\Gamma_3$. Therefore, the necessary and
sufficient condition for single electron pumping is
$|\epsilon_0|<2\min[\Gamma_1,\Gamma_2,\Gamma_3,\Gamma_4]$.

\section{Quantum interference of MFs}
The parity flipping
in the NOT gate discussed above can be understood generally as a quantum
interference of MFs. We consider a loop of $2N$ TSs with a vortex at the
center of the common superconducting substrate. At the initial state, the
odd/even constriction junctions are on/off, with no edge MF in the system.
First, we turn off $\lambda_1$, which produces two edge MFs
$\hat{\gamma}_1$ and $\hat{\gamma}_2$ on TS(1) and TS(2) respectively. By
turning on $\lambda_2$ and $\lambda_{2N}$, and then turning off
$\lambda_3$ and $\lambda_{2N-1}$, and so on so forth, the MF
$\hat{\gamma}_2$ and $\hat{\gamma}_1$ are transported to TS(2l) and
TS(2l+1) respectively. Finally, we turn on $\lambda_{2l}$, which
annihilates the two MFs.

\begin{figure}[t]
\includegraphics[width=6.0cm]{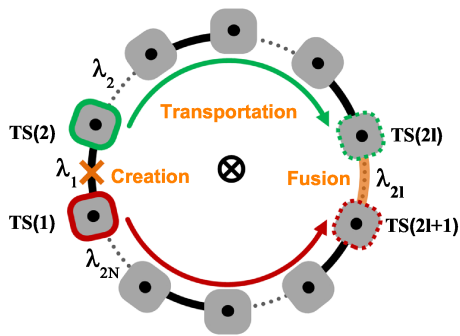} \caption{(Color on line) Interference
of two MFs in a loop of $2N$ TSs with a central vortex. First two MFs are created at TS(1) and TS(2)
by turning off $\lambda_1$, which is denoted by the cross. These two
MFs are then transported to TS(2l) and TS(2l+1) by turning on and off the constriction
junctions on the paths. Finally, they are fused by turning on $\lambda_{2l}$.}
\end{figure}

After the above sequence of switchings, the odd/even constriction
junctions become off/on, opposite to the initial configuration. Let us check
the ground-state electronic parity for the two configurations, keeping in
mind that all the coupling constants $\Gamma_j$ are positive when switched
on as the topological property revealed explicitly for the NOT gate with
four TSs. The ground state $|G\rangle$ at the initial configuration is a
fully occupied state by electrons defined by $\hat{d}_j^{\dag}=
(\hat{\gamma}_{2j-1}+i\hat{\gamma}_{2j})/2$ for $1\le j\le N$. The
electronic parity of the whole system
$\hat{P}=(-1)^{\sum_{j=1,N}\hat{d}^{\dag}_j\hat{d}_j}=\prod_{j=1,N}(-i\hat{\gamma}_{2j}\hat{\gamma}_{2j-1})$
has the eigen value $(-1)^N$.

For the final configuration, the ground state $|\tilde{G}\rangle$ is a
fully occupied state by electrons defined alternatively by $\tilde{d}_j^{\dag}=
(\hat{\gamma}_{2j}+i\hat{\gamma}_{2j+1})/2$ with
$\hat{\gamma}_{2N+1}\equiv \hat{\gamma}_1$. Its electronic parity can be
calculated as
\begin{equation}
\hat{P}|\tilde{G}\rangle=-\prod_{j=1,N} (-i\hat{\gamma}_{2j+1}\hat{\gamma}_{2j})|\tilde{G}\rangle
=(-1)^{N+1}|\tilde{G}\rangle,
\end{equation}
which is opposite to the one of the initial configuration. It is noticed
that when the system of $2N$ TSs is isolated, the parity should be
preserved, and thus the true ground state with reserved parity cannot be
achieved. The attachment of a MF qubit or a QD assists the parity flipping
as discussed above.

It is easy to prove that without the central vortex, or generally with an
even number of vortices, a system of $2N$ TSs takes the same ground-state
parity after switching. Therefore, with an odd number of central vortices,
the quantum interference of two MFs is \textit{constructive}, which flips
the ground-state parity, whereas with an even number of central vortices,
the quantum interference is \textit{destructive} leaving no parity change
in the system. Quantum inference of MFs was discussed for a system with
superconductor and two ferromagnets on the surface of topological
insulator\cite{Akhmerov_Beenakker_prl09}. The two phenomena share the same
topological origin induced by the central vortex. However, in the present
case, the parity is reversed upon the switching process, whereas in the
case addressed before an injected electron at one lead induces either
emission or absorbtion of an electron (strictly speaking, odd number of
electrons), which preserves the parity.

\section{Discussions}
The mechanism of single electron pumping in the present proposal is
different from previous ones in literature \cite{BiasCQ, PWBrouwer,
AdiaCQ}. Unlike the case where electrons are driven by biased potentials
between QD and electrodes\cite{BiasCQ}, we switch the coupling between QD
and the charge reservoir, \textit{i.e.} the NOT gate, instead of the bias. Since the present single
electron pumping is independent of the details of the switching process as
far as it is slow enough, it is also different from the previous adiabatic
pumping schemes where the details of pumping parameters during the
operations have a significant effect on the direction and magnitude of the
pumping current\cite{PWBrouwer, AdiaCQ}. In this sense, the present scheme
can be called topological charge pumping.

In our schemes, the manipulation of edge MFs is topologically
protected by edge excitation gap which sets the limitation of the working
temperature. For a sample size of
150$\times$150nm$^2$, the edge excitation gap is estimated as
$\sim$0.01$\Delta_0$ with $\Delta_0$ the superconduction pair potential
\cite{OurPaper}. For $\Delta_0$=1meV, the gap is $\sim100mK$, which is not
hard to achieve experimentally in these days. It is worthy to note
that there is a work by Akhmerov showing that even the occupation of
higher in-gap excitations will not violate the topological protection
\cite{AkhmerovPRB10}.

\section{Summary}
In conclusion, we reveal that the
mobility of Majorana fermion at edge of nano topological superconductor
induced by a vortex can be used to implement a quantum NOT gate for
Majorana qubit. The working mechanism for the NOT gate can be understood
as the Aharonov-Bohm interference of two Majorana fermions. Based on this
phenomenon, we formulate a scheme for topological single electron pumping.
Useful applications of these devices in quantum transport and quantum
computation are expected.

\vspace{5mm}
\noindent\textbf{Acknowledgements}
This work was supported by WPI
Initiative on Materials Nanoarchitectonics, MEXT of Japan, and
Grants-in-Aid for Scientific Research (No.22540377), JSPS, and partially
by CREST, JST. Q.F.L. is also supported by NSFC under grants 10904092.


\begin{thebibliography}{99}
\bibitem{vonKlitzing} K. von Klitzing, G. Dorda, and M. Pepper, Phys. Rev.
    Lett. \textbf{45}, 494 (1980); K. von Klitzing, Rev. Mod. Phys.
    \textbf{58}, 519 (1986).
\bibitem{Josephson} B. D. Josephson, Phys. Lett. \textbf{1}, 251 (1962);
    \textit{ibid} Rev. Mod. Phys. \textbf{36}, 216 (1964).
\bibitem{GiblinNatComm} S. Giblin, M.
    Kataoka, J. Fletcher, P. See,T. J. B. M. Janssen, J. P. Griffiths, G.
    A. C. Jones, I. Farrer and D. A. Ritchie, Nat. Commun. \textbf{3}, 930(2012); and
    references therein.
\bibitem{LuNature} W. Lu, Z. Ji, L. Pfeiffer, K. W. West and A. J.
    Rimberg, Nature(London) \textbf{423}, 422 (2003).
\bibitem{FujisawaAPL}T. Fujisawa, T. Hayashi, Y. Hirayama, H. D. Cheong
    and Y. H. Jeong, Appl. Phys. Lett. \textbf{84}, 2343 (2004).
\bibitem{BiasCQ} M. H. Devoret, D. Esteve and C. Urbina, Nature(London)
    \textbf{360}, 547 (1992).
\bibitem{PWBrouwer} P. W. Brouwer, Phys. Rev. B \textbf{58}, 10135 (1998).
\bibitem{AdiaCQ} M. Switkes, C. M. Marcus, K. Campman and A. C. Gossard,
    Science \textbf{183}, 1905 (1999).
\bibitem{QHEpumping} S. H. Simon, Phys. Rev. B \text{61}, 16327(2000).

\bibitem{MooreRead91} G. Moore and N. Read, Nuc. Phys. B \textbf{360}, 362
    (1991).
\bibitem{ReadGreen} N. Read and D. Green, Phys. Rev. B \textbf{61}, 10267
    (2000).
\bibitem{Ivanov} D. A. Ivanov, Phys. Rev. Lett. \textbf{86}, 268 (2001).
\bibitem{Kitaev01} A. Y. Kitaev, Phys. -Usp. \textbf{44}, 131 (2001).
\bibitem{Kitaev03} A. Y. Kitaev, Ann. Phys. \textbf{303}, 2 (2003).
\bibitem{DasSarmaPRL05} S. Das Sarma, M. Freedman and C. Nayak, Phys. Rev.
    Lett. \textbf{94}, 166802 (2005).
\bibitem{superfluid} S. Tewari, S. Das Sarma, C. Nayak, C. -W. Zhang and
    P. Zoller, Phys. Rev. Lett. \textbf{98}, 010506 (2007).
\bibitem{MSatoPRL09} M. Sato, Y. Takahashi and S. Fujimoto, Phys. Rev.
    Lett. \textbf{103}, 020401 (2009).
\bibitem{Tewari_Index} S. Tewari, S. Das Sarma and D. -H. Lee, Phys. Rev.
    Lett. \textbf{99}, 037001 (2007).
\bibitem{FuKanePRL08} L. Fu and C. L. Kane, Phys. Rev. Lett. \textbf{100},
    096407 (2008).
\bibitem{MFreturn} F. Wilczek, Nat. Phys. \textbf{5}, 614 (2009).
\bibitem{Akhmerov_Beenakker_prl09} A. R. Akhmerov, J. Nilsson and C. W. J.
    Beenakker, Phys. Rev. Lett. \textbf{102}, 216404 (2009).
\bibitem{SauDasSarmaPRL10} J. D. Sau, R. M. Lutchyn, S. Tewari and S. Das
    Sarma, Phys. Rev. Lett. \textbf{104}, 040502 (2010).
\bibitem{LutchynDasSarma} R. M. Lutchyn, J. D. Sau and S. Das Sarma, Phy.
    Rev. Lett. \textbf{105}, 077001 (2010)
\bibitem{OregvonOppen10} Y. Oreg, G. Rafael and F. von Oppen, Phys. Rev.
    Lett. \textbf{105}, 177002 (2010).
\bibitem{FisherNPhys} J. Alicea, Y. Oreg, G. Refael, F. von Oppen and M.
    P. A. Fisher, Nat. Phys. \textbf{7}, 412 (2011).
\bibitem{TQC_RMP} C. Nayak, S. H. Simon, A. Stern, M. Freedman
 and S. Das Sarma, Rev. Mod. Phys. \textbf{80}, 1083 (2008).
\bibitem{OurPaper} Q. -F. Liang, Z. Wang and X. Hu, Europhys. Lett.
    \textbf{99}, 50004 (2012).

\bibitem{JADAMILU} M. Bollh\"{o}fer and Y. Notay, Comp. Phys. Com.
    \textbf{177}, 951 (2007).
\bibitem{tdbdg_1} S. Roche, Phys. Rev. B \textbf{59}, 2284 (1999).
\bibitem{tdbdg_2} Y. L. Loh, S. N. Taraskin and S. R. Elliott, Phys. Rev.
    Lett. \textbf{84}, 2290 (2000).
\bibitem{AkhmerovPRB10} A. R. Akhmerov, Phys. Rev. B, \textbf{82},
    020509(R) (2010).
\end{thebibliography}
\end{document}